\documentclass[twocolumn]{aastex631}

\newcommand{\rubinsim}{\texttt{RubinSim}}

\usepackage{amsmath}

\begin{document}


\title{An Information-Theoretic Metric for Transient Classification and Novelty Detection}

\author{Ouyang Yu-Qian (Rachel)}
\affiliation{Department of Physics, University of Chicago,
Chicago, IL 60637, USA}

\author{Alex I. Malz}
\affiliation{Space Telescope Science Institute,
Baltimore, MD, USA}

\author{Ming Lian}
\affiliation{University of Delaware,
 Newark, DE 19716, USA}
 \affiliation{Lehigh University,
 Bethlehem, PA 18015, USA}

 \author{Shar Daniels}
\affiliation{University of Delaware,
 Newark, DE 19716, USA}

\author{Federica Bianco}
\affiliation{University of Delaware,
 Newark, DE 19716, USA}
\affiliation{Vera C. Rubin Observatory,
Tucson, AZ 85719, USA}

 \author{Mathilda Nilsson}
\affiliation{University of Delaware,
 Newark, DE 19716, USA}





\begin{abstract}

The development of the observing strategy for the Vera C. Rubin Observatory Legacy Survey of Space and Time (LSST) requires a broad optimization across science cases inside and outside of time-domain astronomy. 
We introduce a novel metric for 
transient science with LSST based on information-theoretic cross-entropy.
We demonstrate its utility for distinguishing populations of objects 
and discuss applications for observing strategy / detection pipeline optimization as well as novelty detection and follow-up resource allocation.

\end{abstract}


\keywords{Methods: statistical — Time domain astronomy — Transient sources — Sky surveys — Information theory}



\section{Introduction}


The Legacy Survey of Space and Time (LSST) at the Vera C. Rubin Observatory is a time-domain discovery machine that will repeatedly image $\geq$ 18,000 square degrees of sky to an unprecedented depth over the course of the next ten years. 
Its detailed observing strategy, parameterized by the frequency, duration, and filter (photometric band) of observations on each sky area, impacts the science outcomes of the raw data.
For example, with a five-day cadence, one could get a high-confidence fit on the light curve of a type Ia supernova; while a kilonova, under the same cadence, would have at most two data points throughout its lifespan -- barely enough to trigger an alert of a change from the baseline template for brightness or colors.
Some faster transients, like stellar flares,
might be detectable only within a single night's observations.

The scientific community thus made a coordinated effort to choose an optimal observing strategy balancing the diverse needs from different research areas, resulting in a baseline choice of observing strategy \citep{pstn-056} under which each sky field is observed in two filters with a $\sim30$ minute separation within a night, and for $\sim4\%$ of the instances, this pair of observations are followed by another observation in one of the two filters used for the initial pair (this set of three observations will be later referred to as a \emph{triplet} in this work). 
The process of optimizing the survey strategy was facilitated by the \rubinsim\ (formerly MAF) software infrastructure.\footnote{\url{https://github.com/lsst/rubin_sim}.} 
This infrastructure is essentially a library of metrics that can be automatically evaluated over any proposed observing strategy, allowing a detailed comparison of how various choices of survey parameters impact each science application.
\citep{bianco2022optimization}.\footnote{A collection of papers dedicated to the assessment of different strategic choices on science is published in the Special Issue of the ApJ Supplements \url{https://iopscience.iop.org/journal/0067-0049/page/rubin_cadence}.}

While a plan is in place for the survey strategy at the start of operations \citep{pstn-056}, modifications continue to be considered for at least two reasons, and in both cases \rubinsim\ will be used to decide whether and how to make such adjustments.
First, our understanding of the telescope may change, e.g. the detector efficiency may be better or worse than initially projected and may change over time. 
Second, the scientific community may reprioritize metrics or introduce new ones, e.g. if a new and exciting type of transient is discovered, it might become a high priority to adjust the cadence to better detect them. Alternatively, if the analysis procedure for a science case advances significantly, that objective could become less or more sensitive to various aspects of the observing strategy.

One science use case that merits further investigation is fast transients as a special case of novelty detection.
This case had little influence on the baseline observing strategy decision, not because it is less interesting to the community or agnostic to cadence, but because appropriate \rubinsim\ metrics were scarce  \citep{Bianco_2019, Andreoni2022serendip, li2022preparing} given how novel the science goal is. The frontier of astronomical discovery has been pushed to faster and faster time scales over the course of the past century, such that when LSST was initially conceived, $\sim$day- and week-time scales were minimally explored, while today the discovery space largely shifted to $\sim$hours- or even minutes \citep{Bianco_2019, Andreoni_2021, li2022preparing,  2023PASP..135j5002H, ho2023search, ho2023minutes}.
This paper builds on earlier metric work \citep{Bianco_2019, li2022preparing, Lian_2021} developed for fast and anomalous transients, which leveraged a characterization of transients in color and magnitude-evolution space. 


In this work, we seek to quantify differences between transient populations using the three nightly measurements available under a given LSST observing strategy.
Following \citet{2018AJ....156...35M, 2020ApJ...890...74K, 2021arXiv210408229M, 2025A&A...694A.130M}, we consult information theory to identify a promising direction for a metric that could be evaluated on-the-fly on incoming data, facilitate the identification of classes of transients, and make recommendations for observing strategy adjustments in subsequent years of LSST and deployed as a tool to facilitate discovery of novel transients.

This paper is organized as follows.
Section~\ref{sec:metric} provides the derivation of the metric.
Section~\ref{sec:data} describes the mock data on which we validate the metric.
Section~\ref{sec:implementation} outlines the specifics of the implementation of our metric in the context of this specific validation data.
Section~\ref{sec:results} presents the results of the demonstration of the metric on the mock data.
Section~\ref{sec:discussion} explores applications of the metric.
We conclude in Section~\ref{sec:conclusion}.

\section{Analysis Metric: Cross-entropy}
\label{sec:metric}

It is desirable to be able to evaluate the probability that a newly-observed transient belongs to a well-known class or to a rare one, or even to an entirely new one \citep{li2022preparing}. This is even more important when transients have fast evolutionary time scales, and therefore recognizing them promptly can enable follow-up. \citep{Bianco_2019}.
The metric proposed in this work is based on the cross-entropy, which quantifies how unexpected a set of observations is under a given probabilistic model.
In this case, the given model is that of transients that are well-understood in the space of Rubin observables, which includes explosive, accreting, and eruptive transients typically with evolutionary timescales of weeks to months, and the new set of observations may be of one or more newly-observed transients of unknown classes. 
A demonstration of the metric on simple Gaussian toy models is provided to help build intuition for the metric in Appendix~\ref{section: toy model}.

\subsection{Theoretical Framework of the Metric}

In information theory, the information content associated with an outcome $x$ occurring with probability $p(x)$ is defined as
\begin{eqnarray}
            \label{eqn:self-information}
                I(x) = -\log_b p(x)
            \end{eqnarray}
in which the base of the logarithm $b>1$ is a scaling factor that determines the units of information;
throughout this work, we use the natural logarithm, corresponding to information measured in nats.

The proposed method operates on discrete distributions defined on a common finite support, consistent with the structure of our data. 
The entropy of a discrete random variable $X$ with probability mass function $p(x)$ is the expected value of the information content,
\begin{eqnarray}
    \label{eqn:entropy}
        H(X) = \mathbb{E}[I(X)]
        = -\sum_{x\in X}p(x)\log p(x) ,
    \end{eqnarray}
which measures the intrinsic uncertainty of the distribution.

Entropy, therefore, provides a natural measure of how surprising it is to observe particular transient events, making it well-suited to anomaly detection. 
However, our goal in this paper is not to quantify the uncertainty of the observed data alone, but rather to measure the relative unexpectedness of newly detected transients compared to the distribution of known events already present in the database. 
For this reason, we apply the cross-entropy, which evaluates the expected surprisal of samples drawn from one distribution in comparison to another.

The cross-entropy between two probability distributions $p$ and $q$ defined over a common support $X$ is given by
\begin{eqnarray}
    \label{eqn:cross entropy 1}
        H(p,q) = H(p) + D_{KL}(p || q) ,
\end{eqnarray}
where $D_{KL}(p || q)$ is the Kullback-Leibler Divergence. 
Equivalently, cross-entropy may be written as
\begin{eqnarray}
    \label{eqn:cross entropy 2}
        H(p,q) = -\sum_{x\in X} p(x)\log q(x) ,
\end{eqnarray}
which directly quantifies the expected surprisal of observations drawn from $p$ under the model $q$. 
Large cross-entropy values correspond to events that are poorly explained by the empirical distribution of known transients, i.e. observations that would be exceptionally rare if drawn from the reference distribution.
In the context of time-domain astronomy, we interpret a large cross-entropy as a signature of novelty, events that would not be expected under the reference distribution.

\subsection{Probability Mass Function of the Metric}
\label{sec: PMF of the metric}

In this work, probability distributions are defined over a discrete phase space $X$ corresponding to binned observable differences derived from triplets of photometric measurements. 
Each outcome $x\in X$ corresponds to a discrete cell in phase space. 

We define a probability mass function $p(x)$ such that
\begin{equation}
     \label{eqn:probability mass function}
    \sum_{x\in X} p(x) = 1 ,
\end{equation}
where $p(x)$ represents the probability that a randomly selected observational triplet occupies phase-space cell $x$. 
Under this definition, cross-entropy directly measures the expected surprisal of observational triplets drawn from one transient population when evaluated under the empirical distribution of another.

This global normalization over the full discretized phase space is a mathematical requirement of the cross-entropy metric and is independent of how the empirical distributions are estimated in practice. 
The specific construction of $p(x)$ from observational data products is described in subsequent sections.

\section{Data: Presto-Color Data Products for Transient Analysis}
\label{sec:data}


The metric we develop is calculated from distributions of different transient types over a space of observable dimensions.
This data product is constructed via a three-step process outlined below.

\subsection{PLAsTiCC light curves under an LSST cadence}
\label{sec:plasticc}

The Photometric LSST Astronomical Time-series Classification Challenge (PLAsTiCC) was an open data challenge aiming to catalyze development of light curve classifiers for the scale and complexity of LSST \citep{2018arXiv181000001T}.
It entailed simulating many instances of several types of transients using the proposed LSST observing strategy at that time to build up a large set of realistically complex three-year light curves for the community to experiment with \citep{2019PASP..131i4501K}.
Both real and synthetic light curves $m_{i, F; C, j}(T_{i})$ are sequences of per-band $F$ magnitudes $m$ at times $T_{i}$ of each transient $j$ of each class $C$.
This rich data set has since been used in dozens of publications studying the observability and utility of transient and variable light curves from LSST.

Among the outcomes of PLAsTiCC were substantiated concerns that the baseline observing strategy would be suboptimal for many time-domain science applications, including fast transients. Under the LSST umbrella, several surveys will be conducted \citep{LPM-17}. This paper only considers the Wide Fast Deep (WFD) LSST program, which will dominate the survey strategy requiring $\sim70-90\%$ of the observing time (depending on yet-to-be-determined survey speed and efficiency).
The LSST observing strategy, both the proposal at the time of PLAsTiCC and the eventual recommendation for early science operations, entails two or three visits at each pointing for each night it is observed: Two photometric filters $(F_{1}$ and $F_{2})$, separated by about $\Delta T_{1}\sim30$ minutes, are always observed, and in about 4\% of cases there will be an additional visit in $F_{1}$ or $F_{2}$ after $\Delta T_2\sim4$ hours the same night.
The pairing of observations is primarily aimed at facilitating tracking of solar system objects \citep{LPM-17}, but the choice of using different filters and the addition of a third observation in a subset of the pairs responds to a desire to enhance transient discovery \citep{Bianco_2019}. The filter choice depends on the illumination (moon phase), filter balance, and sky conditions, and is ultimately determined by the automated LSST scheduler that attempts to optimize observations in the context of the 10-year LSST plan \citep{PSTN-007}.


\subsection{Presto-Color Probability Hypercubes}
\label{sec:prestocolor}
    
\citet{Lian_2021}  reprocessed the PLAsTiCC light curves as part of an effort to quantify the observability of fast transients of types not included in PLAsTiCC, yielding a four-dimensional data cube per object type using the process below.
First, the light curves were linearly interpolated and separated by the true object class $C$.
Observations were grouped by night and object identity into triplets taken in a single night. 
Those triplets represented magnitudes at times $t$, $t+\Delta T_{1}$, and $t+\Delta T_{2}$, where the first and last would be in the same photometric filter and the middle one would be different.\footnote{That is: the first filter in the pair is always the repeat filter from which we derived the magnitude evolution $\Delta m$, i.e. $(F_1,F_2)$ indicates a series of three observations with: $F_1$ at t=0, $F_2$ at t=0+$\Delta T_1$ and $F_1$ at t=0+$\Delta T_2$. By generating the cube including positive and negative values for $\Delta T_1$ and $\Delta T_2$ we implement the method without loss of generality.}
From those magnitudes, a color $c_{1,2} = m_{F_1}(t)-m_{F_2}(t+\Delta T_1)$ and magnitude change $\Delta m_{1,2} = m_{F_1}(t)-m_{F_1}(t+\Delta T_2)$ were calculated.
Thus the PLAsTiCC data were reduced to $\{(F_{1}, F_{2}), (\Delta T_{1}, \Delta T_{2}), (\Delta m_{1,2}, c_{1,2})\}_{C}$ nested quadruples divided only by type of object.
These quadruples were summarized into a discretized four-dimensional histogram with 28 bi-directional band pairs covering all combinations of $ugrizy$ (except $(u,y)$ and $(y,u)$, since these filters are never mounted together on the Rubin Observatory filter wheel, see \citealt{pstn-056});\footnote{in practice, only adjacent or near adjacent filters, (e.g. 
$g$ and $r$).} all pairwise combinations of 64 15-minute bins $-480\, \mathrm{m} \leq \Delta T_{1} \leq 480 \,\mathrm{m}$ 
(which corresponds to an 8 hour span between observations, i.e. a full observing night) and 128 30-minute bins in the range 
($-1920 \,\mathrm{m} \leq \Delta T_{2} \leq -1440 \,\mathrm{m} \mid -480 \,\mathrm{m} \leq \Delta T_{2} \leq 480\,\mathrm{m} \mid 1440\, \mathrm{m} \leq \Delta T_{2} \leq 1920 \,\mathrm{m} $); 
110 0.1-magnitude width magnitude change bins $-5.05 \leq \Delta m \leq 5.95$, and 38 0.5-magnitude width color bins $-9.25 \leq c \leq 9.75$. 
There is one such hypercube for every available class, each summing to the total number of nights of observations over all objects of that class. 





\begin{figure*}
    \includegraphics[width = 1\linewidth]{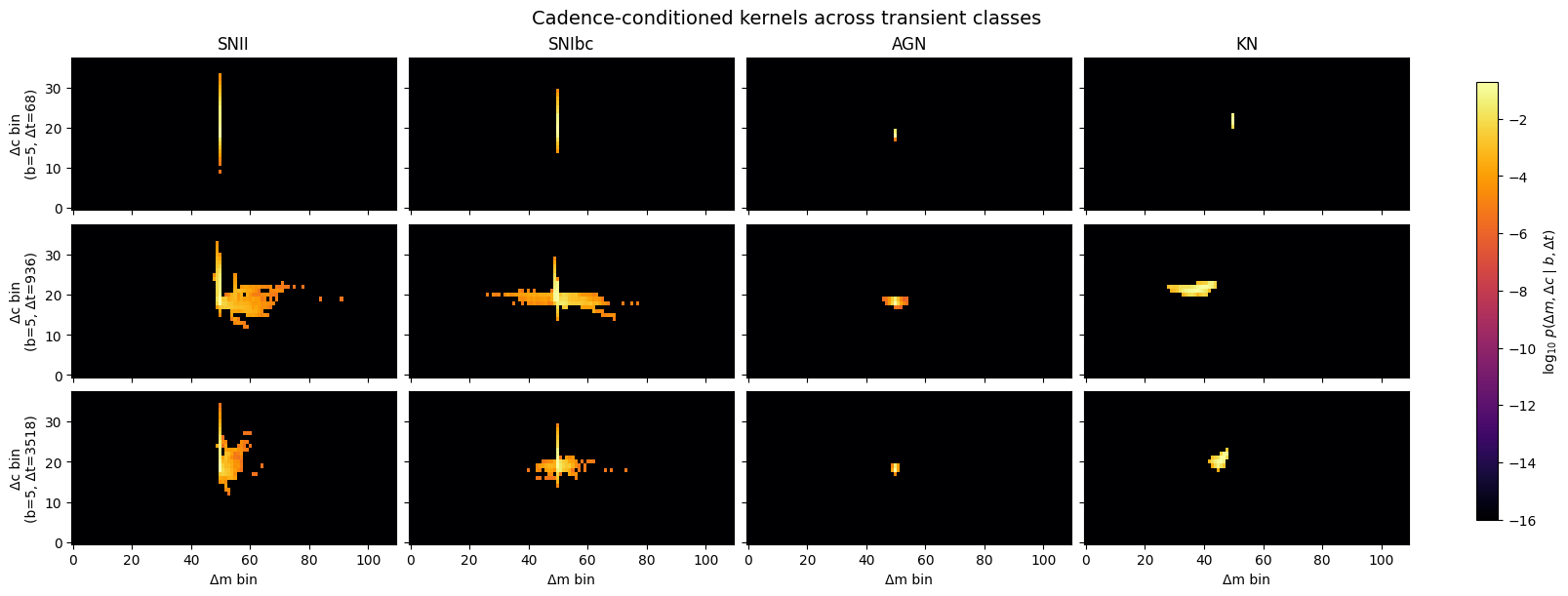}
    \caption{Cadence-conditioned reference kernels for four transient classes (SNII, SNIbc, AGN, KN) at a fixed band pair $b=5$, corresponding to the $g-r$ color and $\Delta m = \Delta g$ magnitude evolution measurements for different time pairs indexed by $\Delta t = 68,\, 936,\, 3518$ (corresponding to the time pairs [-465\,min, 0\,min], [-210\,min, -1470\,min], [465\,min, -480\,min] respectively). We note that $\Delta T_2$=0 correspond to two simultaneous observations in the same filter, i.e. the objects are characterized by color only in the top row. We make this choice since comparing across multiple time pairs (as shown in publicly available code \href{https://github.com/Rachel-0420/Cross-Entropy-Fast-Transient/blob/real-data/cross_entropy_btw_transients.ipynb}{Presto-Color Hypercubes GitHub} line 18), these three show the most visible change within each transient class. Each panel shows the conditional probability mass function $P(i,j \mid C, b=5, \Delta t)$, with color indicating $\log_{10}$ probability. The figure highlights consistent differences in the distribution of effective photometric behavior between transient classes across distinct time separations.}
    \label{fig:kernel 1}
\end{figure*}

\begin{figure*}
    \includegraphics[width = 1\linewidth]{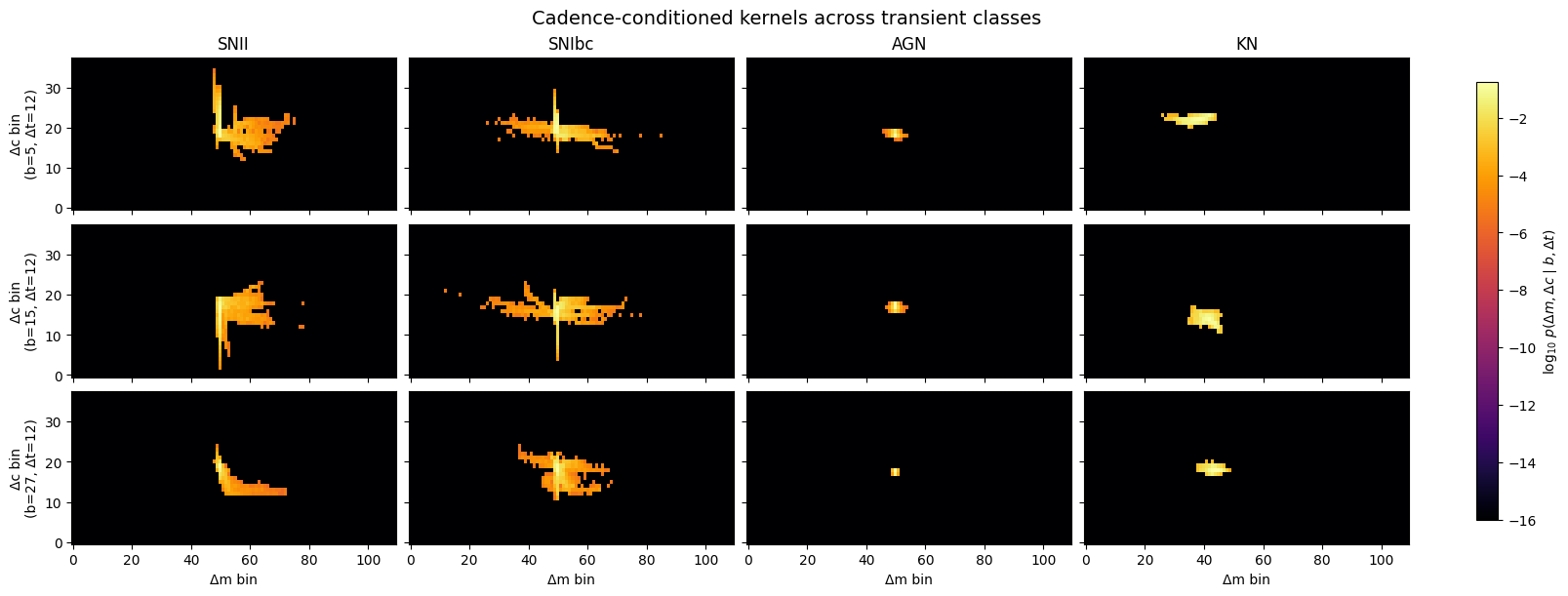}
    \caption{Cadence-conditioned reference kernels for four transient classes (SNII, SNIbc, AGN, KN) at a fixed time pair indexed by $\Delta t=12$ (corresponding to the time pair [-480\,min, -1560\,min]) for different band pairs $b = 5, 15, 27$, corresponding to ($g-r$; $\Delta m = \Delta g$), ($i-g$; $\Delta i$), and ($y-z$; $\Delta y$) triplets respectively. 
    Each panel shows the conditional probability mass function $P(i,j \mid C, b, \Delta t=12)$, with color indicating $\log_{10}$ probability. 
    The figure highlights consistent differences in photometric behavior between transient classes across distinct filter pairs.}
    \label{fig:kernel 2}
\end{figure*}

In practice, the hypercube is indexed by (filter-pair, time-pair, $\Delta m$, color), treating $(F_1,F_2)$ and $(\Delta T_1,\Delta T_2)$ as discrete axes and resulting in a four-dimensional array. 
The raw hypercube is constructed as a histogram of counts from simulated observations: for each filter-pair $b$, time-pair $\Delta t$, and bin $(i,j)$ in the fixed $(\Delta m, c)$ grid, the entry $n_{b,\Delta t,i,j}$ records the number of simulated triplets falling into that bin. 
At this stage, the hypercube contains relative occupancies but is not yet normalized to represent probabilities. 
In subsequent analysis, these counts are normalized to construct empirical probability mass functions over the discrete bin grid, enabling direct statistical comparison between transient classes. Representative slices of these normalized distributions are shown in Figures~\ref{fig:kernel 1} and~\ref{fig:kernel 2}.



\section{Implementation}
\label{sec:implementation}

In this section, we describe how the information–theoretic metric introduced in Section 2 is evaluated in practice using the Presto-Color data products described in Section 3. 
We emphasize that this section concerns operational choices required to apply the metric to our discretized observational data.

\subsection{Cadence-Conditional Reference Kernels}

While cross-entropy is defined on globally normalized distributions over the full four-dimensional phase space, it is useful to factor these distributions into cadence-dependent and photometric components determined by the experimental design of the observations and intrinsic physical behavior of the transient, respectively. 
In the following, we introduce cadence-conditioned kernels corresponding to empirical conditional probability mass functions (PMF) of the reference transient class $C$. 
This factorization clarifies how the joint probability structure informs both population-level comparisons (via cross-entropy) and event-level evaluation (via cadence-conditioned kernels).

As described in Section 3.2, each transient population is represented by a four-dimensional Presto-Color probability hypercube of histogram counts indexed by $(b, \Delta t, \Delta m, c)$, where $b$ denotes the filter pair  $(F_1,F_2)$, $\Delta t$ the time-pair configuration $(\Delta T_1, \Delta T_2)$, and $i,j$ index fixed bins in $(\Delta m, c)$.
The raw hypercube stores the number of simulated observations falling into each bin $n_{b,\Delta t, i, j}$.

For each $(b,\Delta t)$ slice, we locally normalize the counts over the $(\Delta m, c)$ bin grid to obtain a discrete empirical conditional PMF:
\begin{equation}
    \label{eqn:conditional distribution}
    P(i,j\mid C, b, \Delta t) = \frac{n_{b,\Delta t, i, j}}{\sum_{i',j'} n_{b,\Delta t, i', j'}} .
\end{equation}
This normalization ensures that for fixed observing configuration $(b,\Delta t)$, 
\begin{equation}
    \label{eqn:check of conditional distribution}
    \sum_{i,j} P(i,j\mid C, b, \Delta t) = 1 .
\end{equation}
We emphasize that these distributions are defined on a finite discrete phase space determined by the binning scheme; 
we do not approximate a continuous probability density, and bin-area factors do not enter the normalization.
This approximation is made without loss of generality due to the shared binning across all classes in our data set.

For the purposes of evaluating individual observations, we treat each such conditional PMF as a cadence-conditioned reference kernel,
\begin{equation}
    \label{eqn:kernel definition}
    K_{C, b, \Delta t}(i,j) = P(i,j\mid C,b,\Delta t) .
\end{equation}
These kernels represent the expected photometric behavior of known transient populations under a fixed observing cadence and serve as the reference models against which new observations are compared. 
No additional smoothing or parametric modeling is applied; 
the kernels are purely empirical and inherit the binning and normalization of the underlying Presto-Color data products.

We choose four representative classes of astrophysical transients ($C$) as modeled in the PLAsTiCC challenge (see \citealt{kessler2019models}) to demonstrate the use of the Presto-Color probability hypercubes: 
Type II supernovae (SNII), Type Ib/c supernovae (SNIbc), active galactic nuclei (AGN), and Kilonovae  (KN, modeled as in \citealt{bulla2019origin}). We explain this choice in Section~\ref{sec:results}.

Figures~\ref{fig:kernel 1} and~\ref{fig:kernel 2} illustrate representative cadence-conditioned kernels across transient classes. 
For fixed observing configurations, the empirical PMFs exhibit distinct geometric structures in the $(\Delta m,c)$ plane. 
SNII kernels display broader, asymmetric regions of phase space reflecting large-amplitude and color-evolving behavior, while SNIbc distributions are comparatively more localized and elongated along a dominant direction. 
In contrast, AGN kernels remain sharply concentrated within a compact region near the origin, consistent with lower-amplitude stochastic variability, while the KN population is localized in a compact but offset region of phase space, consistent with its characteristically rapid evolution and extreme colors.
Although SNII and SNIbc differ in detailed structure, both occupy comparably extended regions of phase space than those of AGN and KN. 
These qualitative patterns persist across band pairs and time separations, indicating that the cadence-conditioned structure captures stable class-specific photometric signatures stemming from the underlying physical mechanisms of the time-domain phenomena.

\subsection{Scoring Individual Observations}

Given an observational pair $x=(i,j)$, corresponding to a measured $(\Delta m, c)$ falling into the bin $(i,j)$ under cadence $(b,\Delta t)$, we define a surprisal score
\begin{equation}
    \label{eqn:score define}
    s(x\mid C, b, \Delta t)=-\log{K_{C, b, \Delta t}(i,j)} ,
\end{equation}
where $K_{C, b, \Delta t}(i,j)$ is defined in Eq~\ref{eqn:kernel definition}. 

This quantity represents the information content of the observation under the reference transient class $C$. 
Low scores indicate observations that are well explained by the reference population under the given cadence, while large scores correspond to observations that are rare or unexpected within that population. 
In practice, zero-occupancy bins in the empirical kernels are assigned a small floor probability to ensure numerical stability when evaluating the logarithm; this choice is discussed in Appendix~\ref{sec:app-comp}.


\subsection{Global Normalization for Cross-Entropy Evaluation}
\label{sec: global normalization}

The cross-entropy metric defined in Eq~\ref{eqn:cross entropy 2} may be interpreted as the expectation value of the pointwise surprisal derived from the cadence-conditioned kernels, averaged over the full joint distribution of observing cadence and photometric outcomes.
To evaluate the cross-entropy, we construct empirical probability distributions from the Presto-Color probability hypercubes.

As discussed in Section \ref{sec: PMF of the metric}, for the purposes of cross-entropy evaluation, we require a single probability mass function defined over the full discretized space. We therefore perform an additional global normalization step. 

Let $A(x)$ denote the value stored in the hypercube at index $x$, where $x=(b,\Delta t, i,j)$ ranges over all discrete combinations of cadence, color, and magnitude-change bins. 
We define the globally normalized empirical PMF
\begin{eqnarray}
    \label{eqn:empirical probability mass function}
    p(x) = \frac{A(x)}{\sum_{x'\in X}A(x')} ,
\end{eqnarray}
where $X$ denotes the full set of hypercube indices. 
Under this normalization, $p(x)$ represents the probability that a randomly selected observational pair from the transient population has the corresponding band pair, time separation, magnitude change, and color.



Normalized hypercubes are constructed independently for each transient class using the same binning and normalization procedure. 
In the cross-entropy calculation, $p(x)$ denotes the distribution of observational triplets for the examined transient population, while $q(x)$ denotes the corresponding distribution constructed from a known transient population in the database.


Similar to the surprisal score per pair, to ensure numerical stability, we apply a small positive floor to the zero-occupancy bins of the reference distribution prior to global normalization. 

\subsection{Cross-Entropy Evaluation}

The resulting normalized distributions are then used directly to compute pairwise cross entropies according to Eq~\ref{eqn:cross entropy 2} by summing the pointwise contributions $-p(x)\log q(x)$ over all hypercube entries, yielding a scalar measure of how well the observational behavior of the examined transient population is explained by the reference distribution. We emphasize that this formulation compares class-conditional photometric distributions and does not incorporate population-level event rates. Incorporating such rates would enable a fully probabilistic anomaly score and is left for future work.


\section{Validation on Known Transient Classes}
\label{sec:results}

We validate the cross-entropy metric  on the four classes of astrophysical transients introduced in Section \ref{sec:implementation}.
SNII and SNIbc are distinct classes of Core-Collapse supernovae \citep{Arcavi2016, Pian2016}. Their evolution arises from the collapse of a massive star when the stellar core has burn into $Fe^{56}$ \citep{Foglizzo2016}, for a hydrogen-rich, and a hydrogen-poor stellar progenitor, respectively. While distinguishable spectroscopically as well as by the details of their light curve morphology \citep{Gal-Yam2017}, their common origin produces overall similar observable properties. Meanwhile, AGN variability \citep{ulrich1997variability}, produced primarily by changes in the accretion flow onto a supermassive black hole in distant galaxies, is qualitatively distinct; and the population of KNe \citep{metzger2020kilonovae}, modeled after KN170817 \citep{abbott2017phrvl} and understood to arise from the merger of two neutron stars and characterized by rapid neutron capture ($r$-process) nucleosynthesis \citep{barnes2022signatures}, exhibits more distinctly rapidly evolving photometric behavior, making this set well suited for assessing whether the metric captures meaningful population differences.


For each transient class, observational data are drawn from the Presto-Color database and represented as four-dimensional probability hypercubes indexed by filter pair, time separation, magnitude change, and color. 
As described in Section \ref{sec: global normalization}
, each hypercube is globally normalized to define an empirical probability mass function over the full discretized phase space. 
These distributions are constructed independently for each class using identical binning and normalization procedures.




Pairwise cross entropies are then computed between all class combinations according to Eq~\ref{eqn:cross entropy 2}, yielding a scalar measure of how well the observational behavior of one population is explained by the empirical distribution of another. 

\begin{figure}
    \centering
    \includegraphics[width = 0.8\linewidth]{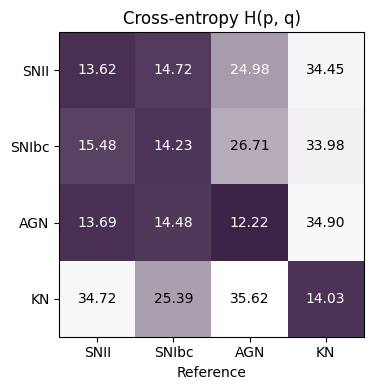}
    \caption{Pairwise cross entropies $H(p,q)$ computed globally between Presto-Color distributions for four transient populations. 
    Each entry represents the expected surprisal of observations drawn from population $p$ when encoded using the distribution of population $q$. 
    Diagonal terms correspond to self cross entropies, while off-diagonal terms quantify population-level dissimilarity. 
    The matrix indicates that SNII and SNIbc exhibit strong mutual similarity, as expected from their related photometric evolution. Although AGN variability arises from a fundamentally different physical mechanism, its cadence-conditioned photometric distribution occupies a compact region that lies within or near regions of non-negligible support in the supernova distributions, resulting in comparatively lower cross-entropy values. In contrast, the kilonova population (KN) is remarkably dissimilar from all other classes, with consistently high cross-entropies, reflecting its localization in a distinct region of observable phase space with minimal overlap under the adopted cadence (see Figure~\ref{fig:kernel 1} and \ref{fig:kernel 2}).
    Note that due to the asymmetry of $D_{KL}(p||q)$, the matrix is not symmetric and the relevant surprisal scores for a transient population $C_X$ should be read along the $C_X$ row.}
    \label{fig:cross entropy}
\end{figure}

Figure \ref{fig:cross entropy} shows the resulting cross-entropy matrix for the three transient populations. 
The diagonal terms correspond to the entropies $H(p)$ of each population. Since $D_{KL}(p||q)\geq 0$,\footnote{For a derivation of this property known as Gibb's inequality, see \citet{breheny_gibbs_2024}.}, from Eq~\ref{eqn:cross entropy 1} follows that one must have $H(p,q) \geq H(p)$. Hence, each row achieves its minimum along the diagonal. Note that no analogous column-wise constraint exists due to the asymmetry of cross-entropy.

As expected, for each population $p$, the smallest value in its row occurs when it is evaluated against its own empirical distribution. The differing diagonal values reflect the intrinsic dispersion of each population in the Presto-Color phase space: more diffuse distributions exhibit larger entropies. 
More particularly, the cross-entropy between SNII and SNIbc is modestly larger than their self-cross entropies, reflecting their related but non-identical photometric behavior. 
In contrast, cross entropies involving AGN are larger, consistent with the physically different variability patterns exhibited by AGN relative to supernova populations, while those involving the kilonova population KN are consistently highest, reflecting its localization in a distinct region of observable phase space with minimal overlap with the other classes (see Figure~\ref{fig:kernel 1} and \ref{fig:kernel 2}).

These results demonstrate that the cross-entropy metric behaves in a physically interpretable manner, assigning lower surprisal to observational triplets drawn from similar transient populations and higher surprisal to those drawn from distinct populations. 
This provides validation that this quantitative metric captures meaningful qualitative structure in the Presto-Color phase space and motivates its use as a hard diagnostic for transient novelty and population differentiation.

Because cross-entropy is asymmetric, i.e. $H(p,q)\neq H(q,p)$, the choice of reference population matters for interpretation. 
The quantity $H(p,q)$ measures how surprising observation triplets drawn from the examined population $p(x)$ would be under the reference distribution $q(x)$. 
A small value indicates that the observed transient might fall within the same class with the reference transient, while a larger value indicates dissonance. 
This asymmetry makes cross-entropy particularly suited to assessing novelty relative to known populations, rather than to defining a symmetric notion of distance.

\section{Discussion}
\label{sec:discussion}

This versatile metric has multiple applications within transient science.
We focus here on two of these applications, namely observing strategy optimization for fast transients and anomaly detection.

\subsection{Interpreting Cross-Entropy as an Observation Strategy Diagnostic}

The cross-entropy provides a quantitative way to assess how the observational behavior of a newly-detected transient compares to that of known transient classes. 
Large cross-entropy values indicate that the observed triplets differ substantially from those typically produced by the reference population, while smaller values indicate closer similarity. 
A toy prototype of this proposed application is provided in Appendix~\ref{sec:cadence}.

Importantly, the comparison is carried out over distributions of observational triplets rather than over individual transients or physical parameters. Each distribution reflects how frequently different combinations of band pair, time separations, magnitude evolution, and color occur under a given observing plan. 
As a result, the cross-entropy value can be understood as measuring similarity given observing pattern, not solely similarity in intrinsic physical properties. 

Two transient populations may yield similar cross-entropy values if they produce comparable patterns of observation triplets, even if their underlying physical mechanics and even photometric behavior differ.
Consider a very coarse observing pattern focusing on certain photometric filters under which a pair of classes do not differ significantly, neglecting other photometric filters over which that pair of classes exhibit significant differences.
In this scenario, the cross-entropy may be small, whereas the cross-entropy under an observing strategy with more balanced coverage in photometric filters could be larger.

In this way, the cross-entropy could serve as a metric for observing strategy optimization (the original scope of \citealt{Bianco_2019}), a process that will continue throughout the LSST's ten-year mission.
Such metrics must be repeatedly evaluated as the high-dimensional space of observing strategy parameters is explored and new potential observations of each class are simulated.
The computational simplicity of the cross-entropy relative to that simulation process makes it an apt choice for such a metric.
Moreover, it can be applied to all transient and variable classes, enabling an apples-to-apples comparison;
one could not only make statements along the lines of, ``this observing strategy is worse for science with classes X and Y because it would reduce the ability to distinguish them observationally'' but also say how that information difference compares to that between classes Z and W that may be more distinguishable under that same observation strategy. This application of our information-theory metric is discussed in detail in \ref{sec:cadence}.

It is worth noting that the Rubin Observatory plan for science-driven survey optimization \citep{bianco2022optimization} and the construction of the Metric Analysis Framework to enable it and facilitate it \citep{jones2014lsst} were initially conceived \citep{https://doi.org/10.5281/zenodo.842713} to support information-theoretic metrics \citep[like][]{2026ApJS..282...69M} that would measure the bits of information added by a set of observations to the data that supports a science case. These metrics proved to be extremely hard to define for most science cases, and the community effort largely moved to producing simulation-based metrics, where populations of astrophysical phenomena are introduced into an observing strategy and the quality of retrieved data measures the value of a survey strategy \citep{2022ApJS..258....3R, 2022ApJS..258....4H, 2022ApJS..258....5A, 2022ApJS..258...23A, 
2022ApJS..259...58L,
2025ApJS..281....6B,2025ApJS..276...10A,2024ApJS..273...35D,2023ApJS..268...13B,2023ApJS..267...15S,2023ApJS..266...22S,2023ApJS..265...43A,2023ApJS..265...41D,2023ApJS..265...27B,2023ApJS..264...22G,2022ApJS..260...18A}. This, however, makes the relative gain/loss of performance across different science cases impossible to compare formally (this aggregation is currently handled by the Survey Cadence Optimization Committee on the basis of their scientific expertise and in consultation with the scientific community at large). 
The cross-entropy would thus be an improvement over the current collection of inhomogeneous per-class metrics that cannot be compared in such a uniform manner.


\subsection{Implications for Unsupervised Classification and Decision-making Applications}

The cross-entropy formulation also provides a natural mechanism for reference selection. 
Observations of a candidate transient may be evaluated against multiple reference distributions, and the effective reference class determined by the minimum cross-entropy. 
In this sense, the metric not only quantifies divergence from a given population, but also implicitly identifies the most compatible reference model. 
In other words, it could be used to define and distinguish new populations from established populations, for example, for identifying novel subtypes whose observables are embedded within the range of those of their parent class.
It could even serve as an appropriate internal objective function for unsupervised classifier development.
This flexibility is particularly useful in survey settings where the appropriate comparison baseline may not be known in advance. 
Even in cases where the transient represents an anomaly, its pattern of relative cross-entropies across known classes provides information about both its proximity to certain populations and its separation from others, thereby aiding in the characterization of ensembles of new events.
In other works \citep{Bianco_2019, Lian_2021}, we had highlighted the use of the Presto-Color hypercubes to rapidly identify unusual objects as they fall on low-occupancy cells and demonstrated that even a single triplet of observations in two filters can flag an unusual object when placed in a Presto-Color hypercube. However, here the focus is on population-level methods: the cross-entropy is a population-level metric and it is particularly effective at  at empirically identifying separate classes or subclasses, particularly when the evolutionary time-scales are significantly different and assessing the ability of a survey strategy to separate different classes (e.g. KNe from SNe).

The potential application to classification also suggests applications to decision-making for other analysis choices such as transient detection algorithm and follow-up observation strategy.
Consider plotting the observables (either in projection or under a dimensionality reduction algorithm such as t-SNE or UMAP ---\citealt{hinton2002stochastic, Hinton03, Maaten08, McInnes18}) of potential candidates for spectroscopic follow-up observations over those of known populations.
Visually, for exploring novel classes, we'd want to follow up objects that have poor overlap with the known classes;
similarly, for optimally following up rare objects of a known class, we'd seek to select those that overlap well with the area defined by the previously identified objects of that class.
The cross-entropy gives us a way to quantify what we would otherwise attempt to do by eye, making it possible to select objects automatically.








\section{Conclusion}
\label{sec:conclusion}
In this work, we introduce an information-theoretic metric for quantifying differences between transient populations in discretized photometric phase space. 
By representing each transient class as an empirical probability distribution over cadence-conditioned observables and evaluating the cross-entropy between these distributions, we obtain a scalar measure of population-level dissimilarity. 
This construction provides a principled tool for quantifying the impact of changes in the observing strategy as well as for classifying newly detected transients into known categories and identifying novel anomalous populations. 

Using Presto-Color probability hypercubes constructed from PLAsTiCC light curves under an LSST-like observing strategy, we demonstrated that the metric captures systematic differences between well-characterized transient classes. 
In particular, cross-entropy comparisons reflect known similarities between SNII and SNIbc populations and their qualitative distinction from AGN variability, while also highlighting the stronger separation of the kilonova population (based on models by \citealt{bulla2019origin}), which occupies a more localized and distinct region of observable phase space. These results indicate that the metric responds to meaningful structure in the joint distribution of magnitude change and color, conditioned on the separations in time between the observations.

A key feature of this framework is that it operates directly on empirical distributions conditioned on observing cadence, enabling the metric to automatically incorporate survey-specific sampling effects. 
Because the method requires only discretized observational products, it is computationally tractable and can be evaluated on-the-fly for newly detected events as well as on synthetically modeled data.

We further discuss potential applications to unsupervised classification and decision-making, suggesting future applications of the method. 
We also explore extensions of the method to cadence optimization (see Appendix~\ref{sec:cadence}). 
The study of pair-wise cross-entropy, rather than only aggregated measures specific to each science case, encodes the sensitivity of different cadences to class-level distinctions with a scalar in common units.
Comparing pair-wise cross-entropy on data of known transient classes therefore can be used to derive guidance on matters of cadence-selection and survey strategy design.

While the present study focuses on a limited set of transient classes and simulated light curves, the approach is general and can be extended to additional populations, alternative aggregation schemes, or higher-dimensional observational parameters.

\appendix
\twocolumngrid
\section{Demonstration on Toy Models}
\label{section: toy model}

To illustrate the behavior of the cross-entropy metric in a controlled setting, we apply it to a set of simple two-dimensional toy distributions with analytically defined structure. 
This experiment serves as a sanity check for the discrete implementation described in Section~\ref{sec:implementation} and clarifies how the metric responds to structured differences in location and covariance.
It is our aim for these examples to impart intuition to the reader regarding the behavior of the metric relative to more traditional metrics.

\begin{figure}
    \centering
    \includegraphics[width=1\linewidth]{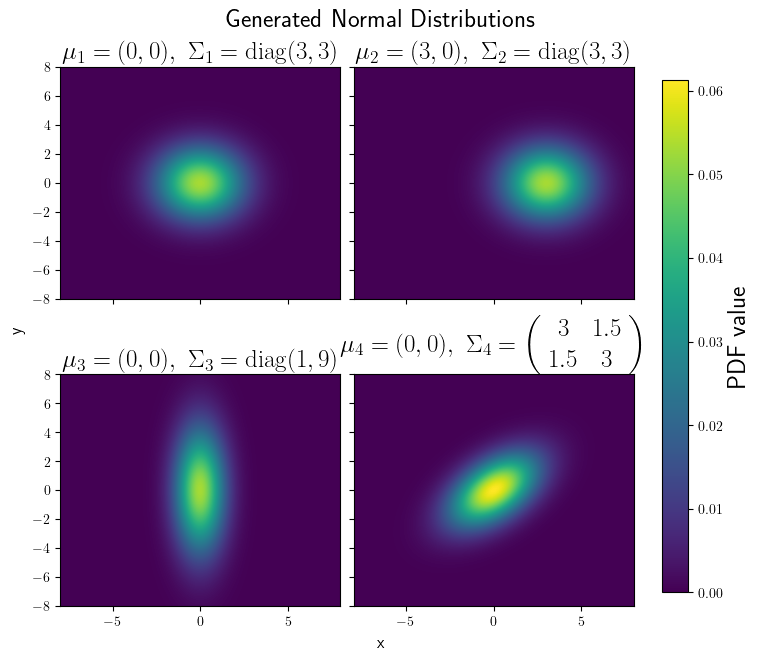}
    \caption{Gaussian reference distributions $p_1, p_2, p_3, p_4$ (upper-left, upper-right, lower-left, lower-right). 
    Each panel shows a two-dimensional marginal of a multivariate normal distribution. 
    The standard reference $p_1$ has mean $\mu = (0,0)$ and covariance $\Sigma=\mathrm{diag}(3,3)$. 
    Distribution $p_2$ shifts the mean relative to $p_1$ while keeping the covariance fixed. 
    Distribution $p_3$ modifies the covariance to an anisotropic diagonal form without shifting the mean.
    And $p_4$ introduces nonzero off-diagonal covariance, inducing correlation and rotating the principal axes of the distribution while preserving the mean.}
    \label{fig:normal distribution}
\end{figure}

Let us consider four bivariate normal distributions $p_i(x)$, $i=1,2,3,4$, defined over $x \in \mathbb{R}^2$ with means $\mu_i$ and covariance matrices $\Sigma_i$, 
\begin{equation}
    \label{eqn:simple distribution}
    p_i(x) = \mathcal{N}(x \mid \mu_i, \Sigma_i) .
\end{equation}
The distributions are chosen by distinct types of variation, covering mean-shift, anisotropy, and correlation. 
Figure \ref{fig:normal distribution} visualizes the marginal structure of these distributions, with $p_1,\, p_2, \, p_3, \, p_4$ corresponding to the four types respectively.

\subsection{Cross-Entropy in the Continuous Case}

For continuous distributions $p(x)$ and $q(x)$, cross-entropy is defined as \begin{equation}
    \label{eqn:cts cross entropy}
    H(p,q) = -\int p(x)\,\log q(x)\,dx .
\end{equation}
When both $p$ and $q$ are multivariate normal, one can derive a closed-form expression for the cross-entropy:
\begin{equation}
    \label{eqn: closed-form expression, cts normal cross entropy}
    \begin{split}
    H(p,q)
    = \frac{1}{2} \Big[
    &\log \det (2\pi \Sigma_q) \\
    &+ \mathrm{tr}(\Sigma_q^{-1}\Sigma_p) \\
    &+ (\mu_q - \mu_p)^\top
      \Sigma_q^{-1}
      (\mu_q - \mu_p)
    \Big] .
    \end{split}
\end{equation}
This expression separates the contributions from differences in mean and covariance, allowing direct interpretation of how distributional changes influence the cross-entropy metric. 

\begin{figure}
    \centering
    \includegraphics[width=0.8\linewidth]{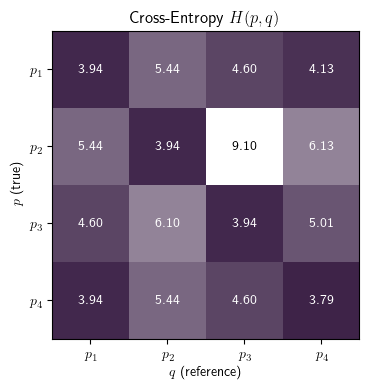}
    \caption{Continuous cross-entropy matrix $H(p,q)$ computed using the closed-form expression of Eq~\ref{eqn: closed-form expression, cts normal cross entropy}. 
    Rows correspond to the true distribution $p$, and the columns correspond to the reference distribution $q$. 
    Diagonal elements represent the entropies $H(p)$ of each distribution, while off-diagonal terms quantify dissimilarity between distributions. 
    Larger values indicate greater divergence in mean or covariance structure. The dominant entropy occurs for the pair $(p_2,p_3)$, reflecting that a mean-shifted Gaussian is poorly described by an anisotropic Gaussian with a different covariance structure, while distributions sharing similar structure exhibit comparatively smaller cross-entropies.}
    \label{fig:cts normal cross entropy}
\end{figure}

Since $p_i(x)$ are initially generated as continuous multivariate normal distributions, we can directly apply Eq~\ref{eqn: closed-form expression, cts normal cross entropy} for cross-entropy evaluations. 
The result is shown in Figure \ref{fig:cts normal cross entropy}. 

\subsection{Cross-Entropy in the Discrete Case}

To mirror the procedure applied to the Presto-Color hypercubes, we convert each continuous Gaussian density into a discrete probability mass function by integrating the density over a fixed rectangular grid of bins. 
The resulting bin probabilities are then renormalized to sum to unity on the finite domain, yielding a discrete PMF on which cross-entropy is evaluated.

Let $f(x,y) = \mathcal{N}\big((x,y)\mid \mu,\Sigma\big)$ denote a bivariate normal density on $\mathbb{R}^2$. 
We introduce a rectangular partition of the plane with bin edges $\{x_i\}$ and $\{y_j\}$ covering a bounded region chosen to capture essentially all of the Gaussian mass. 
For each bin $[x_i, x_{i+1}] \times [y_j, y_{j+1}]$, we assign probability mass by integrating the continuous density over the bin,
\begin{equation}
    p_{ij} = \int_{x_i}^{x_{i+1}} \int_{y_j}^{y_{j+1}} f(x,y)\,dy\,dx .
\end{equation}

Because the Gaussian probability density has units of inverse area, this integration removes units and produces dimensionless probabilities. 
The resulting collection $\{p_{ij}\}$ defines a discrete distribution on the grid. 
Since the Gaussian has infinite support, the finite domain captures slightly less than total probability one; 
to ensure the sum of all $\tilde p_{ij}$ is unity, we therefore renormalize as
\begin{equation}
    \tilde p_{ij} = \frac{p_{ij}}{\sum_{k,\ell} p_{k\ell}} .
\end{equation}

\begin{figure}
    \centering
    \includegraphics[width=0.8\linewidth]{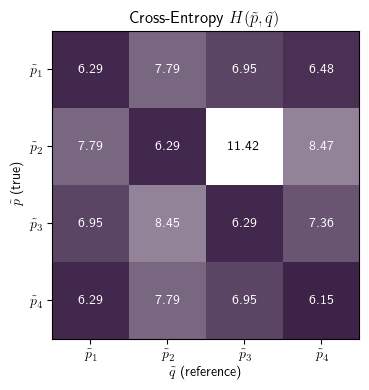}
    \caption{Discrete cross-entropy matrix obtained after projecting each Gaussian distribution onto a finite grid and renormalizing bin probabilities. 
    Rows correspond to the true distribution $P$, and columns correspond to the reference distribution $Q$. 
    Cross-entropy is evaluated using the discrete form Eq~\ref{eqn:discrete cross entropy form}. 
    Differences relative to Figure 5 arise from finite-domain truncation and discretization effects.}
    \label{fig:discrete normal cross entropy}
\end{figure}

Cross-entropy is then evaluated in discrete form,
\begin{equation}
    \label{eqn:discrete cross entropy form}
    H(\tilde p, \tilde q) = -\sum_{i,j}\tilde p_{ij}\log \tilde q_{ij} ,
\end{equation}
which approximates the continuous expression Eq~\ref{eqn: closed-form expression, cts normal cross entropy} and serves as a direct analogy to the methodology applied to transient distributions in the main analysis. 
In this construction, bins are treated as categorical states, and no differential (continuous) entropy is computed.
We apply the discussed procedure, and the resulting cross-entropy matrix is shown in Figure~\ref{fig:discrete normal cross entropy}.

\subsection{Interpretation}

Figure~\ref{fig:cts normal cross entropy} shows the cross-entropy matrix computed from the closed-form expression for continuous multivariate Gaussians, while Figure~\ref{fig:discrete normal cross entropy} displays the corresponding matrix obtained after discretizing each Gaussian onto a finite grid and evaluating cross-entropy using the discrete PMF formulation.
The absolute values differ between the two panels because the discretization introduces a finite bin width and a global renormalization on the bounded domain, resulting in an additive offset relative to the continuous quantity.

Importantly, however, the qualitative structure of the matrix is preserved. 
In both panels, the diagonal elements are smallest, reflecting self-consistency of each distribution. 
The largest off-diagonal entries occur for the pair $(P_2,P_3)$, indicating that a mean-shifted Gaussian is poorly described by an anisotropic Gaussian with different covariance structure. 
Distributions that differ only by moderate covariance deformation exhibit intermediate cross-entropy values. 

These results confirm that the metric responds systematically to structured distributional differences. 
In particular, it increases with separation in location and with divergence in covariance;
this is because entropy-based metrics tend to be sensitive to mutual coverage, and little probability mass is shared when the separation exceeds the covariance, whereas changes in the covariance, even complex ones, have a smaller impact on mutual coverage. 
This toy-model experiment therefore provides an interpretable demonstration of the metric’s behavior before application to the more complex observational distributions considered in the main text.

\subsection{Pairwise KL Divergence: Toy Model for a Cadence-Information Score}
\label{sec:cadence}

The object of this subsection is to construct a simplified prototype of the optimization principle proposed in the main text for cadence selection.

Whereas we have been mirroring the four Gaussian distributions to different transient classes, we may also let them mirror the distribution of observables for one transient class under different cadences. 
In the full analysis, each cadence configuration $c$ induces a class-conditional distribution $P_{i}^{(c)}$. 
The pairwise KL divergence matrix therefore becomes cadence-dependent,
\begin{eqnarray}
    \label{eqn:KL divergence through cross-entropy matrix}
    \begin{split}
        D_{ij}^{(c)} &= D_{KL}(P_i^{(c)}|| P_j^{(c)})\\
        &= H(P_i^{(c)}, P_j^{(c)}) - H(P_i^{(c)}, P_i^{(c)}) 
    \end{split} .
\end{eqnarray}

\begin{figure}
    \centering
    \includegraphics[width=0.8\linewidth]{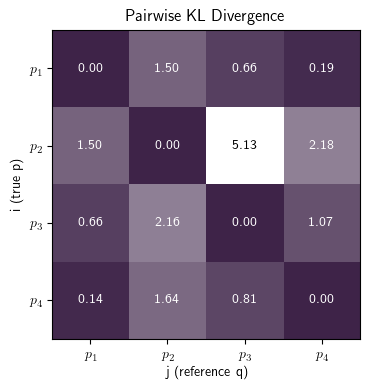}
    \caption{Pairwise KL divergence matrix for the discretized Gaussian distributions. 
    Each entry is computed through Eq~\ref{eqn:KL divergence through cross-entropy matrix} and represents the expected log-likelihood ratio in favor of $p_i$ over $p_j$. 
    Larger off-diagonal values indicate greater statistical distinguishability between distributions. 
    As in Figures \ref{fig:cts normal cross entropy} and \ref{fig:discrete normal cross entropy}, the dominant divergence occurs for the pair $(p_2,p_3)$, while distributions sharing similar structure exhibit comparatively smaller divergences.}
    \label{fig:pairwise KL}
\end{figure}

Applying this to the discretized cross-entropy matrix, the result is shown in Figure~\ref{fig:pairwise KL}. 
Each off-diagonal entry quantifies the statistical distinguishability between a pair of distributions. 
However, cadence optimization requires a single scalar objective rather than a full pairwise matrix.

To convert the matrix into a cadence-level criterion, we aggregate the pair-wise divergence into a global score. 
The guiding principle is that a cadence is more informative if it increases the distinguishability between class-conditioned distributions. 
For KL divergence, larger values correspond to greater statistical distinguishability. 
Hence, we define a symmetric aggregate information score by averaging over the asymmetric divergences across unordered pairs:
\begin{equation}
        S(c) = \frac{2}{N(N-1)}\sum_{i<j}\frac{D_{ij}^{(c)} + D_{ji}^{(c)}}{2} .
    \label{eqn:symmetric aggregate score}
\end{equation}

Maximizing $S(c)$ over cadence configurations selects the cadence that maximizes average statistical distinguishability between classes. 
This has three desirable properties: 
\begin{enumerate}
    \item It compresses a full matrix into a single scalar suitable for optimization.
    \item It removes the asymmetry of KL divergence, which makes sense for classes with no a priori relationships, e.g. hierarchical structure.
    \item Each pair of classes contributes equally to the aggregated score, though weights could be introduced to deliberately favor or disfavor classes based on scientific prioritization (as in \citealt{2019AJ....158..171M}).
\end{enumerate}


Although the present experiment involves simple Gaussian distributions, it mirrors the structure of the observational transients: 
structured differences in mean and covariance translate directly into differences in KL divergence, and hence into differences in the aggregate information score. 
The toy model thus provides an interpretable demonstration of the optimization mechanism.

\section{Computational Details}
\label{sec:app-comp}

Prior to global normalization, all hypercube arrays are converted to floating-point format to ensure numerical stability. 
Zero-occupancy bins in reference distributions are assigned a small positive floor value $10^{-16}$ before normalization in order to avoid undefined logarithms in the cross-entropy computation. 
Conditional normalization over $(\Delta m, c)$ bins is verified explicitly for each $(b, \Delta t)$ slice, and global normalization is checked to ensure that the resulting empirical probability mass functions sum to unity over the full discretized phase space.

All probability normalizations and cross-entropy calculations are implemented in publicly available code: 
\href{https://github.com/Rachel-0420/Cross-Entropy-Fast-Transient/blob/real-data/cross_entropy_btw_transients.ipynb}{Presto-Color Hypercubes GitHub} and \href{https://github.com/Rachel-0420/Cross-Entropy-Fast-Transient/blob/real-data/HighD%20Normal%20Distribution.ipynb}{Toy Model GitHub}.
The repository includes unit tests that confirm both conditional and global normalization constraints. Code to produce the Presto-Color hypercubes can be found at \href{https://github.com/lmptc/PrestoColor2.git}{PrestoColor2}.

\section*{Acknowledgments}
\begin{acknowledgments}
We thank Rachel Mandelbaum for supervision at the onset of this work. 
During the course of this work, A.I.M. was supported by Schmidt Sciences. F.B.B. acknowledges the support of NSF award 2511639. S.D. is supported by a NSF Graduate Student Fellowship: this material is based upon work supported by the National Science Foundation Graduate Research Fellowship Program under Grant No. 2444111. Any opinions, findings, and conclusions or recommendations expressed in this material are those of the author(s) and do not necessarily reflect the views of the National Science Foundation.
\end{acknowledgments}

\bibliography{bibliography, lsst}{}
\bibliographystyle{aasjournal}



\end{document}